\RequirePackage{fix-cm}
\documentclass[twocolumn]{svjour3}          
\smartqed  

\usepackage{graphicx}
\usepackage{subfig}
\usepackage{times}
\usepackage{amsmath}
\usepackage{amssymb}
\usepackage{etoolbox}
\usepackage{multirow}
\usepackage{xcolor}
\usepackage{array}
\usepackage{slashbox}
\usepackage[export]{adjustbox}
\usepackage{mathtools}
\usepackage{arydshln}
\usepackage{setspace}
\usepackage{mathptmx}
\DeclarePairedDelimiter{\ceil}{\lceil}{\rceil}
\begin{document}

\title{ASDN: A Deep Convolutional Network for Arbitrary Scale Image Super-Resolution
}

\titlerunning{ASDN}        

\author{Jialiang Shen        \and
        Yucheng Wang \and
        Jian Zhang 
}

\authorrunning{Jialiang Shen, Yucheng Wang, and Jian Zhang} 

\institute{Jialiang Shen \at
              University of Technology Sydney \\
              Tel.: +61-420-390-016\\
              \email{Jialiang.Shen@student.uts.edu.au}             \\
           \and
           Yucheng Wang \at
              Bytedance\\
              Tel.: +86-136-8102-4248\\
              \email{wangyucheng.ai@bytedance.com}  \\
           \and
           Jian Zhang \at
              University of Technology Sydney\\
              \email{Jian.Zhang@uts.edu.au}  \\
}


\maketitle

\begin{abstract}
Deep convolutional neural networks have significantly improved the peak signal-to-noise ratio of Super-Resolution (SR). However, image viewer applications commonly allow users to zoom the images to arbitrary magnification scales, thus far imposing a large number of required training scales at a tremendous computational cost.
To obtain a more computationally efficient model for arbitrary-scale SR, this paper employs a Laplacian pyramid method to reconstruct any-scale high-resolution (HR) images using the high-frequency image details in a Laplacian Frequency Representation.
For SR of small-scales (between 1 and 2), images are constructed by interpolation from a sparse set of precalculated Laplacian pyramid levels. SR of larger scales is computed by recursion from small scales, which significantly reduces the computational cost.
For a full comparison, fixed- and any-scale experiments are conducted using various benchmarks.  At fixed scales, ASDN outperforms predefined upsampling methods (e.g., SRCNN, VDSR, DRRN) by about 1 dB in PSNR. At any-scale, ASDN generally exceeds Meta-SR on many scales.

\end{abstract}

\section{Introduction}
\label{intro}
Deep neural networks have made good progress in Single-image Super-Resolution (SISR), adeptly extracting image priors from data sets and efficiently learning mapping functions from LR to HR patches. However, for applications that allow users to zoom to arbitrary scales (e.g., face image SR~\cite{gao2018face} and satellite image SR~\cite{lu2019satellite}), multi-scale methods which learn the LR to HR mapping functions independently at each of several scales~\cite{lim2017enhanced}\cite{zhang2018residual}\cite{kim2016accurate} become inefficient. Meta-SR \cite{hu2019meta} shows that SR of arbitrary decimal scales can be achieved by training one single model with the dynamic meta-upscaling module. But meta-SR can only generate HR images on scales for which it has trained, making it computationally impractical to train for all scales of interest for any-scale SR.

To alleviate the need for so many training scales, we find image patches have the same similarity at different scales. The self-similarity-based SR method~\cite{lu2017underwater} enhances the textural content with similar patches across different scales. Furthermore, image edges are scalable, and different-scale images have similar edge information, represented by high-frequency image information. In order to seek the missing high-frequency information of SR images, a Laplacian pyramid based-method is proposed to interpolate between a sparse set of trained scales. Indeed, the Laplacian filter is an edge detector, and the Laplacian noise term can be used to detect the outliners for robust tracking~\cite{wang2015robust}. Therefore, similar high-frequency image information across different scales can be highlighted through the Laplacian pyramid structure. Moreover, the Laplacian pyramid structure has been proved to reduce the training data requirements for multi-scale SR in MS-LapSRN~\cite{lai2018fast}, generating the $3\times$ HR images with the $4\times$ SR results and predicting $8\times$ HR images by progressively deploying through the network for $2\times$ SR. Therefore, it is feasible to reduce the training costs with a Laplacian Pyramid~\cite{lai2017deep} network structure.

Unlike previous Laplacian Pyramid networks for multi-scale SR, we seek to train a model to predict any-scale SR images. Obviously, a large upsampling ratio can be expressed as an integer power of ratios in a small range. Therefore, given a network for super-resolution at scales in a small range (such as the real-number interval $(1,2]$), arbitrary larger scales (real numbers greater than 2) can be implemented by recursion. Inspired by the classical Laplacian pyramid method~\cite{burt1987laplacian}, which reconstructs HR images by restoring the residual images between two Laplacian pyramid levels, we introduce a Laplacian Frequency Representation to learn the mapping function for SR of scales in the small range $(1,2]$. Our algorithm represents the HR images of any continuous decimal scale in the range by the two neighboring Laplacian pyramid levels. For SR of the large decimal ratios, we progressively upscale the coarse HR images, and recursively deploy them through the network multiple times with a small decimal ratio in the range to gradually refine the HR images.

In this paper, we propose our network as Any-Scale Deep Super-Resolution Network (ASDN) based on the multi-scale parallel reconstruction architecture. Each reconstruction branch shares the Feature Mapping Branch (FMB) and predicts the Laplacian pyramid levels through the Image Reconstruction Branch (IRB). Our network requires a minimal amount of training data and computational resources but effectively generates any-scale SR results.

We present extensive comparisons on both fixed integer scales and any decimal scale on commonly used benchmarks, and provide the results of the ASDN and the fine-tuned ASDN (FSDN), for the reference in comparison with the existing multi-scale SR methods. ASDN outperforms all of the other predefined upsampling methods and even some single upsampling models, without training on the specific-scale data samples. FSDN has state-of-the-art performance for fixed scale SR, comparing favorably to all existing methods. For any-scale SR factor, we retrain many previous network structures~\cite{lim2017enhanced}\cite{zhang2018residual}\cite{kim2016accurate} with our any-scale SR method into any-scale SR categories for comparison. Our ASDN is effective for SR of any desired scale and specifically achieves the state-of-the-art performance on scales within the small range $(1,2]$.

In summary, our work provides the following contributions:

(1) \textbf{Laplacian Frequency Representation:} We propose a Laplacian frequency representation mechanism to reconstruct image SR at small scales, those continuously varying between 1 and 2. The HR images are the weighted interpolation of their two neighboring Laplacian pyramid levels, which efficiently reduces the training scale demands for learning the SR at continuous scales.

(2) \textbf{Recursive Deployment:} We introduce Recursive Deployment for generating the HR images of the larger upsampling ratios, as we find that the HR images of the larger scales can be gradually upsampled and recursively deployed with small ratios. This extends any-scale SR from small scales to larger ones without requiring additional training scales.

(3) \textbf{Any-scale Deep SR Network:} We propose an Any-Scale Deep Super-Resolution Network (ASDN) to generate HR images of any random scale with one unified network, providing enormous computational savings over directly applying existing CNN-based multi-scale methods for any-scale applications.

\section{Related Works}
\subsection{Image Super-Resolution Using CNN}
\begin{figure*}[t]
\begin{center}
\includegraphics[width=0.90\textwidth,height=0.21\textheight]{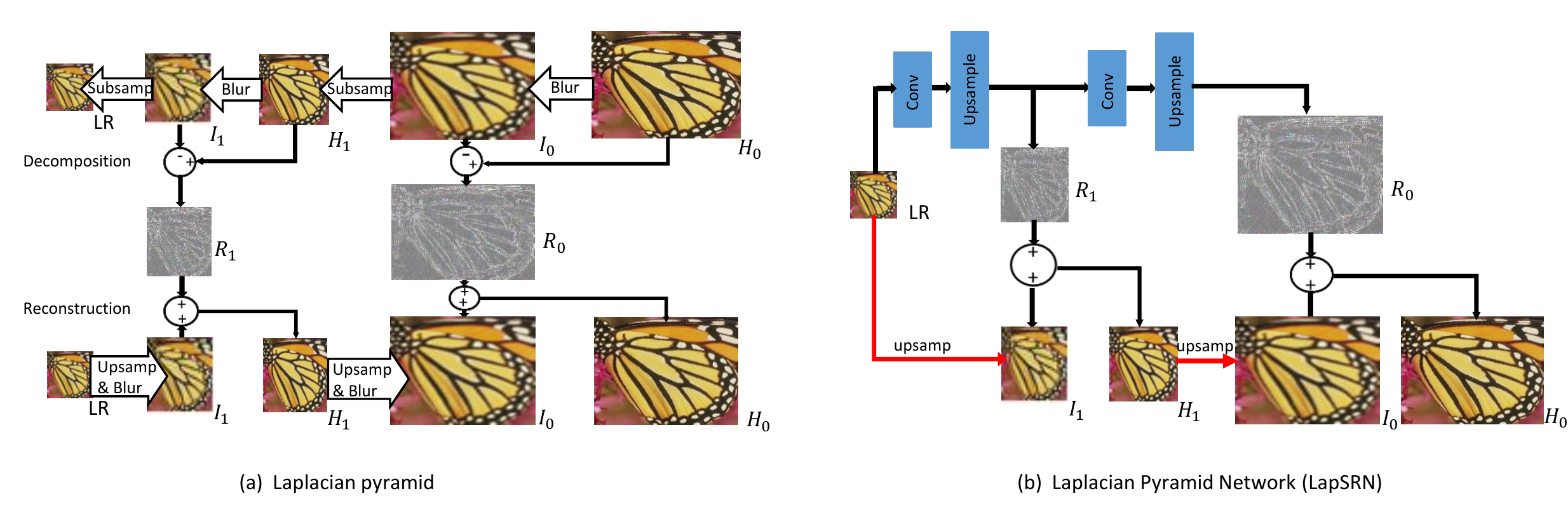}
\caption{\textbf{Comparison of two-level Laplacian Pyramids.} (a) Laplacian Pyramid~\cite{burt1987laplacian}. The decomposition step produces two residual images $R_{1}, R_{0}$ by subtracting $H_{1}$ with $I_{1}$, $H_{0}$ with $I_{0}$ to preserve the high-frequency information, which then added with the interpolated LR $I_{1}, I_{0}$ respectively to reconstruct the $2\times, 4\times$ frequency levels $H_{1}, H_{0}$. (b) LapSRN~\cite{lai2017deep}. The residual images $R_{1}, R_{0}$ are progressively learned by the networks and upsampled at each level, then added with $I_{1}, I_{0}$ for HR images $H_{1}, H_{0}$.}
\label{laplacianfre}
\end{center}
\end{figure*}
Image super-resolution has evolved greatly over the past decades, and numerous image SR methods~\cite{lim2017enhanced}\cite{zhang2018residual}\cite{kim2016accurate} have been proposed to improve image reconstruction performance. With the fast development of the computation processor, CNN-based SR methods have demonstrated state-of-the-art results by optimizing an end-to-end network to learn the LR-HR mapping function. Dong et al. \cite{dong2014learning} initially introduced convolutional layers into image SR, which have been proved effective for the task. However, the network consists of only three layers, unable to observe superior results with the deeper model. He et al. \cite{huang2016deep} solved this problem by residually skip connecting layers inside the network to help the gradient flow across the deeper models. Later on, more skip connection structures, dense connection \cite{tong2017image} were proved to accelerate network convergence by feature reusing across the layers. RDN \cite{zhang2018residual} and DBDN \cite{wang2018deep}, embed the dense convolutional neural network into image SR to further improve image reconstruction accuracy. Then, the attention module was adopted into the SR to help the network focus on the high-frequency feature learning. Liu et al. \cite{liu2018attention} introduced the spatial attention to mask out the high-frequency component locations in the HR images,and RCAN~\cite{zhang2018image} replaced normal feature layers with residual channel layers to adaptively rescale channel-wise features to reduce the unnecessary computations for abundant low-frequency features. However, these methods mainly focus on multi-scale SR (e.g., $2\times, 3\times$, and $4\times$). In this paper, we propose to reconstruct any-scale SR with a few numbers of training scales, which can significantly reduce the computational cost.

Any-scale SR model is seldom investigated in image SR. Recently, Meta-SR \cite{hu2019meta} proposed a meta-upscale module for arbitrary scale SR, which dynamic magnifies image with decimal scales, by training and testing with $40$ different scales at the stride of $0.1$. However, Meta-SR \cite{hu2019meta} did not provide a systemic approach or experimental results for any scale that not included in the $40$ trained scales. In other words, only training with $40$ different scales, Meta-SR can not solve the SR of undetermined decimal scales. Nevertheless, if we use enormous scales of data to train the Meta-SR model for the full any-scale SR approximation, it might take a very long time to optimize the network for its convergence, which is not practical. Different from these methods trained with all the scales of interest, we propose a novel network ASDN for SR of any potential scale, which adopts our any-scale SR method, including Laplacian Frequency Representation and Recursive Deployment.

\subsection{Laplacian Pyramid Structure}

The Laplacian Pyramid~\cite{burt1987laplacian} is used for restoring HR images by preserving residual image information. As shown in Fig.~\ref{laplacianfre}(a), the decomposition step firstly preserves the residual information $R_{1}, R_{0}$, as image downscaled. Then the kept residual information $R_{1}, R_{0}$ will be stored back by adding with the low-resolution image $I_{1}, I_{0}$, to reconstruct the initial HR image $H_{1}, H_{0}$.

With the development of deep learning, many models adopt the Laplacian Pyramid structure as the main mapping frameworks, which construct progressive upsampling networks for image SR. Such as LapSRN~\cite{lai2017deep} in Fig.~\ref{laplacianfre}(b), a multi-phase network and each phase learn the residual information with convolutional layers. LapSRN progressively reconstructs each pyramid levels at the interval of $2$ times, for $2\times, 4\times,$ and $8\times$ SR, respectively. MS-LapSRN~\cite{lai2018fast} is the parameter sharing version of LapSRN, which shares the network parameters across pyramid levels and exhibits the efficiency of recursive deployment. However, these models are designed to effectively predict SR of large scale factors.

In this paper, we present Laplacian Frequency Representation to reconstruct SR results of continuous scales. In our design, each pyramid level is at the interval of $0.1$ in scale and parallelly allocated at the end of the network. According to the Laplacian pyramid~\cite{burt1987laplacian} that the lost high-frequency information can be presented by the two neighboring pyramid levels, the high-frequency information of HR image is predicted based on the weighted interpolation of the two Laplacian pyramid levels neighboring the testing scale. The Laplacian Frequency Representation entails fewer training SR samples to generate HR images of scales in the continuous ratio range, which reduces the undesired training data storage space and shrinks the optimization period to accelerate the network convergence.

\section{Any-Scale Image Super-Resolution} \label{sec:sections}

In this section, we provide the mathematical background of the any-scale SR method including Laplacian Frequency Representation and Recursive Deployment and introduce the structure of our proposed ASDN.
\begin{figure*}[t]
\begin{center}
\includegraphics[width=0.90\textwidth,height=0.20\textheight]{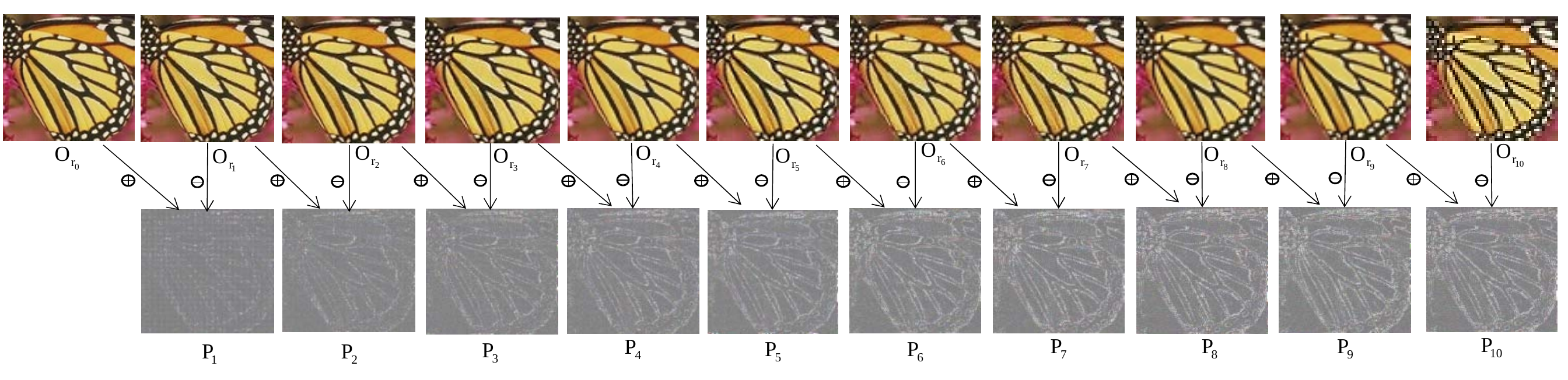}
\caption{The Laplacian Frequency Representation has $11$ Laplacian pyramid levels ($O_{r_{0}}, ..., O_{r_{10}}$), with $10$ phases in the scale range of $(1, 2]$ ($P_{1}, ..., P_{10}$). Each phase represents the difference between the successive Laplacian pyramid levels.}
\label{phase}
\end{center}
\end{figure*}
\subsection{Any-Scale SR Method}
There are two steps in the any-scale SR method: Laplacian Frequency Representation and Recursive Deployment. The proposed Laplacian Frequency Representation method is to generate HR images of decimal scales in a continuous scale range, and Recursive Deployment is to define the recursion times $N$ and the small ratio $r$ at each recursion for any-scale SR prediction. To use the minimum training samples, we define the small decimal ratios $r\in(1,2]$. For SR of each upscaling ratio $R$, the HR image of upscale ratio $R$ can be achieved by recursively upscaled with a small ratio $r$ and deployed $N$ times.

\subsubsection{Laplacian Frequency Representation}
To generate SR of decimal scales in a continuous range, the intuitive method is to train the network with random dense scales in the range. However, we find this method is difficult due to a large amount of training scale samples and computation power for optimizing the network. To deal with this problem, we introduce Laplacian Frequency Representation as the intermediate representation of the high-frequency image information of SR results.

As shown in Fig. \ref{phase}, our proposed Laplacian Frequency Representation has $L$ Laplacian pyramid levels, and each pyramid level $l$ is tasked with learning the high-frequency image information of HR images $O_{r_{l}}$ for the scale $r_{l}$ with training samples of corresponding scales.
\begin{equation}
r_{l}=\frac{l}{L-1}+1 , l=0,...,L-1\frac{}{}
\end{equation}

According to the scalability of the image edges and the comprehensive coverage of high-frequency information of images in edges, we can interpolate the high-frequency image details of SR results of any small decimal scales $r$ based on their two neighboring Laplacian pyramid levels.

For a given scale factor $r$ in this continuous range, the Laplacian frequency represented HR images $O_{r}$ can be defined as
\begin{equation}
O_{r}=O_{r_{i}}+w_{r}*P_{i} \label{2}
\end{equation}
where

\begin{equation}
P_{i}=O_{r_{i-1}}-O_{r_{i}}, i=1,...,L-1 \label{1}
\end{equation}

Here the phase number $i=\ceil{(L-1)*(r-1)}$ and $w_{r}$ is the weight parameter of the edge information for the $r$ scale SR. We define the weight parameter according to distance proportion of the scale $r$ to the $r_{i}$ in the phase $P_{i}$.
\begin{equation}
w_{r}= (L-1)*(r_{i}-r)\label{3}
\end{equation}

The interpolated representation can be regarded as calculating the missing high-frequency image details of HR images of certain scales, so we name the mechanism as Laplacian Frequency Representation. The further evaluation of the accuracy of Laplacian Frequency Representation and the density of Laplacian pyramid levels in the experiment section proves that the represented SR results highly coordinate with the directly learned results, and the performance is stable when the Laplacian pyramid levels are at a certain density.
As a result, we propose to train the Laplacian pyramid levels using deep neural networks with several scales and reconstruct the HR images of continuous decimal scales in the range with Laplacian Frequency Representation.

\subsubsection{Recursive Deployment}
For SR of any upsampling ratio $R$ in the larger range, it is impossible to train SR samples of all the scales to learn the mapping function of these scales. To minimize the training sample demands, we reuse the learned mapping network for SR of decimal scales in the range of $(1,2]$. We are based on the idea that any upscale decimal ratio $R$ can be expressed as an integer $N$ power of decimal ratios $r$ in a small range. Therefore, the HR images of $R$ can be generated by gradually upscaling and recursively deploying through the mapping network $N$ times with small decimal ratios $r\in (1, 2]$.
\begin{figure*}[t]
\begin{center}
\includegraphics[width=0.9\textwidth,height=0.28\textheight]{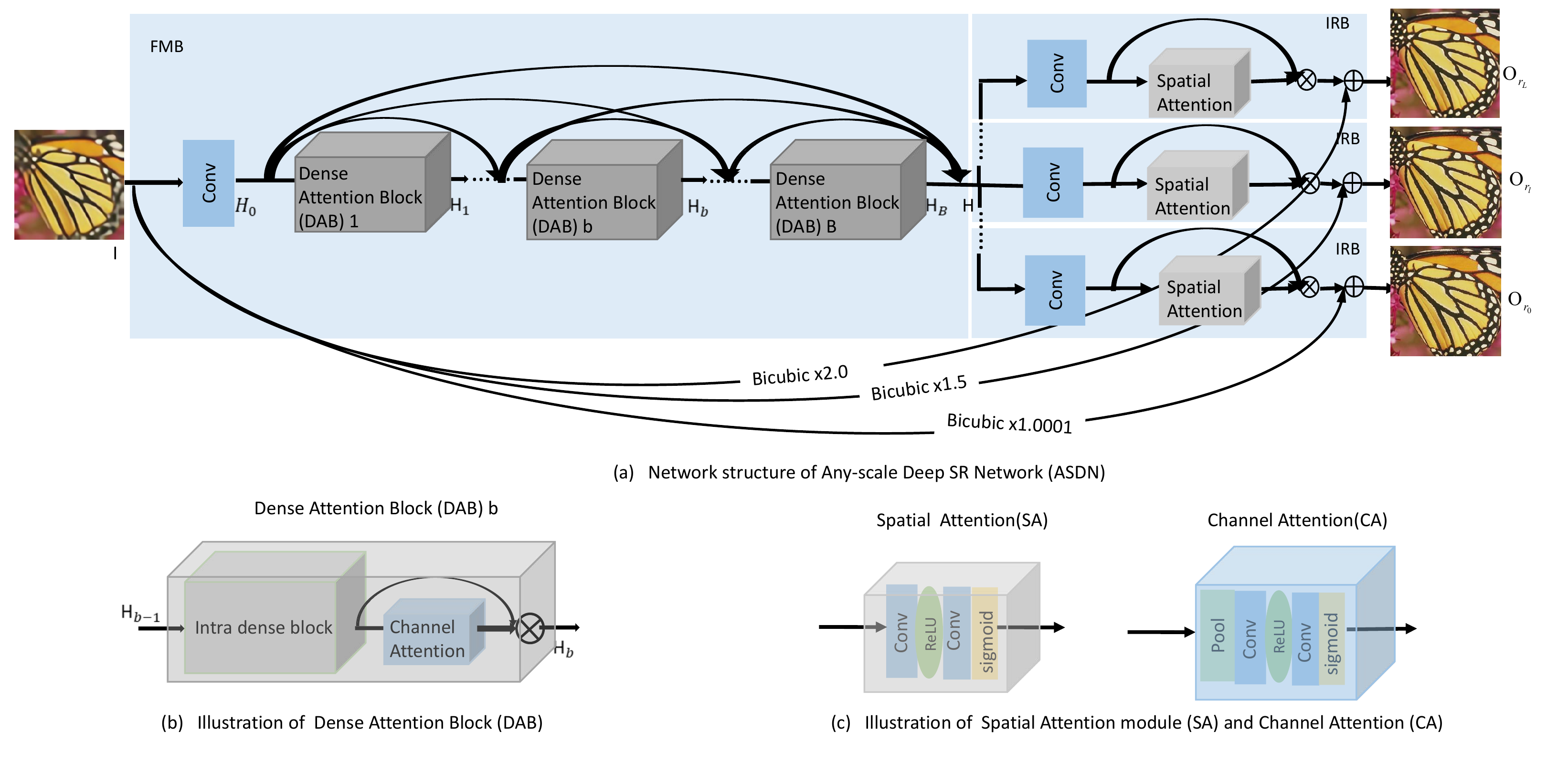}
\caption{(a) The overall architecture of the proposed ASDN network, multiple Image Reconstruction Branches (IRBs) parallelly allocate after Feature Mapping Branch (FMB). The FMB adopts the bi-dense structure from DBDN~\cite{wang2018deep}, and the spatial attention in IRB is the same spatial attention module from CSFM~\cite{hu2019channel}. (b) The dense attention block (DAB) in (a), which combines the Intra dense block from DBDN~\cite{wang2018deep} and channel attention module from RCAN~\cite{zhang2018image}. (c) The illustration of the adopted spatial attention (SA) and channel attention (CA) modules from CSFM and RCAN~\cite{hu2019channel}\cite{zhang2018image}.}
\label{network}
\end{center}
\end{figure*}
We express the $R$ as an integer $N$ power of small decimal ratios $r \in (1, 2]$. The integer $N$ denotes the recursion times for the deployment, and the small ratio $r$ is the upsampling ratio at each recursion. To determine the best solution of $N$ and $r$ for any-scale SR, several comparison experiments are performed in the experiment section. As we observed, SR with the larger upscale ratio $r$ at the early recursions and the smaller recursive deployment times $N$ has better performance than other $N$ and $r$ solutions.

Therefore, for any scale factor $R$, the recursive times $N$
\begin{equation}
N=\ceil{\log_2 R}  \label{4}
\end{equation}
The upscale ratio $r_{n}$ at each recursion $n$ can be defined as
\begin{equation}
  r_{n}=
  \begin{cases}
  2 & \text{if} \,  n \leq N-1 \\
  \frac{R}{2^{N-1}} & \text{if} \,  n= N
  \label{5}
  \end{cases}
\end{equation}
Based on the defined $N$ and $r$ solution for $R$, if the recursion time $N$ is $1$, the HR images of $R=r$ are directly deployed by the network. In other situations, the coarse HR images from the previous recursion are bicubic upscaled with the small ratio $r$ as the input LR images at the current recursion. For better SR performance, at the early $N-1$ recursions, the small ratio $r_{n}=2$, and at the $N_{th}$ recursion, $r_{n}=\frac{R}{2^{N-1}}$.

\subsection{Any-scale SR deep network}\label{anynet}

In this section, we build a deep neural network to predict the Laplacian Frequency Representation from the input images.

\subsubsection{Network Architecture}

The Laplacian Frequency Representation should consist of $L=11$ Laplacian pyramid levels for SR in the scale range $(1, 2]$. Each Laplacian pyramid level is the reconstructed HR image containing high-frequency details. Due to the mutual relationship among different scales in the SR networks~\cite{kim2016accurate}, our network for Laplacian Frequency Representation are based on the multi-scale parallel \cite{zhang2018image} framework by sharing the Feature Mapping Branch (FMB) across different scales and restoring HR images with separate Image Reconstruction Branches (IRBs). Sharing the FMB can largely reduce the computation capacity, and separating IRB reduces the complexity of the original learning problem and leads to an accurate result.

The feature mapping branch (FMB) of the Laplacian Frequency Representation is constructed by a deep convolutional neural network $H = f_{fmb}(I)$. As shown in Fig. \ref{network}, FMB consists of Bi-Dense structure~\cite{wang2018deep} for efficient feature learning and channel attention modules~\cite{zhang2018image} for highlighting high-frequency context information. In the Dense Attention Block (DAB), channel attention module (see Fig. \ref{network}(c)) connects right after the concatenated feature channels. Therefore, the high-frequency information of the concatenated channel features are highlighted before preceding into the next block and thus allow the network to focus on more useful channels to improve reconstruction performance.

The other part of the network $f_{irb}$ is the image reconstruction branch (IRB), which represents the Laplacian pyramid levels. For each Laplacian pyramid level $O_{r_{i}}=f_{irb}(H), i=0,...,10$, the locations of the tiny textures are different, and these textures usually contain high-frequency information, while the smooth areas have more low-frequency information. Therefore, to recover high-frequency details for image SR of different scales, it is helpful to mask out the discriminative high-frequency locations with spatial attention mechanism~\cite{liu2018attention}. As shown in Fig. \ref{network}(a), the learned high-level features are firstly restored into image space by a three-channel convolutional layer at each Laplacian pyramid Level. Then the restored image goes into the spatial attention (SA)~\cite{hu2019channel} unit in Fig. \ref{network}(c), to mask out the adaptive high-frequency information in the HR images of different scales. To preserve the smooth areas information and concentrate on training high-frequency information, the input interpolated LR images are added with the network output by identity skip connection (SC) to generate HR images.

To train the Any-scale SR Deep Network (ASDN) and generate Laplacian Frequency Representation, each IRB is randomly selected and combined after FMB at each update. For some practical applications where only require SR of specific scales, our ASDN can be fine-tuned to a fixed-scale network (FSDN) to further improve the reconstruction accuracy for the scales of interest by training image samples of specific scales. FSDN shares the same network structure as ASDN, except the deconvolutional layer of a specific scale, is inserted at the front of each IRB, which follows the common multi-scale single upscaling SR networks~\cite{lim2017enhanced}\cite{zhang2018residual}\cite{zhang2018image}.

\section{Experiments}
In this section, we describe the implementation details of our models, including model hyper-parameters, training and testing details. Then we compare the proposed any-scale network and the fine-tuned fixed-scale model with several state-of-the-art SR methods on both fixed and any scale benchmark datasets including the quantitative, qualitative comparisons and any-scale comparisons. The effectiveness evaluation of the proposed any-scale method and the contribution study of different components in the proposed any-scale deep network are also provided in the paper.

\subsection{Implementation Details}

\textbf{Network settings} In the proposed ASDN, all convolutional layers have $64$ filters and $3\times3$ kernel size except the layers in IRB for restoring images and the convolutional layers in CA and SA units. The layers for image restoration have 3 filters and all the convolutional layers in CA and SA units are $1\times1$ kernel size, which adopt the same setting as CSFM~\cite{hu2019channel}. Meanwhile, the $3\times3$ kernel size convolutional layer zero-pads the boundaries before applying convolution to keep the size of all feature maps the same as the input of each level. ASDN and FSDN share the same FMB structure, where $16$ DAB are densely connected and each DAB has $8$ dense layers. But in FSDN, the deconvolutional layer settings follow single upsampling networks~\cite{lim2017enhanced}\cite{zhang2018residual} to upscale feature mappings with the corresponding scales. 

\textbf{Training details} The original training images are from DIV2K dataset~\cite{agustsson2017ntire} and Flicker dataset~\cite{agustsson2017ntire}. The input LR images for ASDN are bicubic interpolated from the training images with $11$ decimal ratios $r$, which are evenly-distributed in the range of $(1,2]$. In each training batch, $16$ augmented RGB patches with the size of $48\times48$ are extracted from LR images as the input, and the LR images are randomly selected from one scale training samples among the total $11$ scales training data. Here the data augmentation includes horizontal flips and $90$-degree rotations are randomly adopted on each patch.
To fine-tune the FSDN, the input LR images are downscaled by the scale factor among $2\times, 3\times, 4\times,$ and $8\times$. In the training batch, a batch of $96\times96$ size patches is used as the targets and the corresponding scale LR RGB patches to optimize the specific scale modules.
In general, ASDN and FSDN are all built with the platform Torch and optimized by Adam with L1 loss by setting $\beta_{1}=0.9$, $\beta_{2}=0.999$, and $\epsilon=10^{-8}$.
The learning rate is initially set to $10^{-4}$ and halved at every $2\times10^{5}$ minibatch updates for $10^{6}$ total minibatch updates.
\begin{table*}[t]
\renewcommand{\arraystretch}{0.9}
\tiny
\caption{Quantitative evaluation of state-of-the-art SR algorithms. We report the average PSNR/SSIM for $2\times$, $3\times$, $4\times$ and $8\times$ SR. \color{red}{Red} \color{black}{indicates the best performance, and} \color{blue}{blue} \color{black}{indicates the best performance among predefined upsampling methods.}}
\label{tab3}
\centering
\makebox[1\textwidth][c]{%
\resizebox{0.95\textwidth}{!}{
\begin{tabular}{|c|c|cc|cc|cc|cc|cc|}
\hline
  \multirow{2}{*}{ scale} & \multirow{2}{*}{Algorithms} & \multicolumn{2}{c|}{ Set5}
                                                                          & \multicolumn{2}{c|}{ Set14}
                                                                           & \multicolumn{2}{c|}{ BSD100}
                                                                            & \multicolumn{2}{c|}{ Urban100}
                                                                           & \multicolumn{2}{c|}{ Manga109}\\
                                                                         &  & \textbf{PSNR} & \textbf{SSIM}
                                                                         & \textbf{PSNR} & \textbf{SSIM}
                                                                         & \textbf{PSNR} & \textbf{SSIM}
                                                                          & \textbf{PSNR} & \textbf{SSIM}
                                                                          & \textbf{PSNR} & \textbf{SSIM}  \\

  \hline
  \multirow{12}{*}{$2\times$} &Bicubic & 33.64 & 0.929  & 30.22 & 0.868  & 29.55 & 0.842  & 26.66 & 0.841  & 30.84 & 0.935 \\
  & SRCNN~\cite{dong2014learning} &36.65 & 0.954  & 32.29 & 0.903  & 31.36 & 0.888  & 29.52 & 0.895 & 35.72 & 0.968 \\
  & VDSR~\cite{kim2016accurate} &37.53 & 0.958  & 32.97 & 0.913 & 31.90 & 0.896  & 30.77 & 0.914 & 37.16 & 0.974 \\
  & DRRN~\cite{he2016deep} &37.74 & 0.959 & 33.23 & 0.914  & 32.05 & 0.897  & 31.23 & 0.919 & 37.52 & 0.976 \\
  & LapSRN~\cite{lai2017deep} &37.52 & 0.959  & 33.08 & 0.913  & 31.80 & 0.895  & 30.41 & 0.910  & 37.27 & 0.974 \\
  & MemNet~\cite{tai2017memnet} & 37.78& 0.959& 33.28& 0.914& 32.08 &0.898 & 31.33 & 0.919 & 37.72 & 0.974\\
  & SRMDNF~\cite{zhang2018learning} & 37.79& 0.960& 33.32& 0.916 & 32.05 & 0.899& 31.33& 0.920 & 38.07 & 0.976\\
  & ASDN(ours) & \color{blue}{38.12}& \color{blue}{0.961} & \color{blue}{33.82} & \color{blue}{0.919} &\color{blue}{32.30} &\color{blue}{0.901} & \color{blue}{32.47}& \color{blue}{0.931} &\color{blue}{39.16} &\color{blue}{0.978}\\\cline{2-12}
  & EDSR~\cite{lim2017enhanced} &38.11 &0.960 &33.92 & 0.920 & 32.32& 0.901 & 32.93 & 0.935 & 39.10 & 0.976 \\
  & RDN~\cite{zhang2018residual}&38.24 &0.961 &34.01 & 0.921 & 32.34 & 0.902  & 32.96 & 0.936 & 39.19 &0.978 \\
  & DBPN~\cite{haris2018deep}  &38.09 & 0.961 & 33.85 & 0.920   & 32.27 &0.900 & 32.55 & 0.932 & 38.89 &0.978 \\
  & RCAN \cite{zhang2018image}  &38.27& 0.961 & 34.12 & 0.922&  32.41& 0.903 & \color{red}{33.34} &\color{red}{0.938} &  39.44&0.979 \\
  & FSDN(ours) & \color{red}{38.27}& \color{red}{0.961} & \color{red}{34.18}& \color{red}{0.923} & \color{red}{32.41} &\color{red}{0.903}& 33.13 & 0.937 & \color{red}{39.49} &\color{red}{0.979}\\\hline

  \multirow{12}{*}{\textbf{$3\times$}}&  Bicubic &30.39 & 0.867  & 27.53 & 0.774  & 27.20 & 0.738  & 24.47 & 0.737  & 26.99 & 0.859  \\
  & SRCNN~\cite{dong2014learning}  &32.75 & 0.909 & 29.30 & 0.822  & 28.41 & 0.786& 26.25 & 0.801 & 30.59 & 0.914 \\
  & VDSR~\cite{kim2016accurate} & 33.66 & 0.921  & 29.77 & 0.831  & 28.82 & 0.798  & 27.41 & 0.830  & 32.01& 0.934 \\
  & DRRN~\cite{he2016deep} & 34.03 & 0.924 & 29.96 & 0.835  & 28.95 & 0.800  & 27.53 & 0.764  & 32.42 & 0.939 \\
  & LapSRN~\cite{lai2017deep} & 33.82 & 0.922  & 29.87 & 0.832& 28.82 & 0.798 & 27.07& 0.828 & 32.21 & 0.935  \\
  & MemNet~\cite{tai2017memnet} & 34.09 & 0.925 & 30.00 & 0.835 & 28.96 & 0.800& 27.57 & 0.839 & 32.51 & 0.937\\
  & SRMDNF~\cite{zhang2018learning} & 34.12 & 0.925 & 30.04 & 0.838 & 28.97& 0.802& 27.57& 0.839 & 33.00 & 0.940\\
  & ASDN(ours)  & \color{blue}{34.48} &\color{blue}{0.928} & \color{blue}{30.35} & \color{blue}{0.843} & \color{blue}{29.18} & \color{blue}{0.808} & \color{blue}{28.45} & \color{blue}{0.858} & \color{blue}{33.87} & \color{blue}{0.947} \\\cline{2-12}
  & EDSR~\cite{lim2017enhanced} & 34.65 & 0.928 & 30.52 & 0.846 & 29.25& 0.809 &28.80 & 0.865& 34.17& 0.948 \\
  & RDN~\cite{zhang2018residual} & 34.71 & 0.929& 30.57 & 0.847  & 29.26 & 0.809  &28.80 & 0.865 &34.13& 0.948\\
  & RCAN \cite{zhang2018image}  &  34.74& 0.930& \color{red}{30.65} & 0.848  & 29.32 & 0.811 &\color{red}{29.09} &\color{red}{0.870} & 34.44& 0.949 \\
  & FSDN(ours) & \color{red}{34.75} &\color{red}{0.930} &  30.63 &\color{red}{0.848}  &\color{red}{29.33}&\color{red}{0.811}& 28.98 & 0.868 & \color{red}{34.53}& \color{red}{0.950} \\\hline

 \multirow{12}{*}{\textbf{$4\times$}}& Bicubic & 28.42 & 0.810& 26.10 & 0.704  & 25.96 & 0.669  & 23.15 & 0.660  & 24.92 & 0.789 \\
  & SRCNN \cite{dong2014learning} & 30.49 & 0.862 & 27.61 & 0.754  & 26.91 & 0.712  & 24.53 & 0.724  & 27.66 & 0.858 \\
  & VDSR~\cite{kim2016accurate} & 31.35 & 0.882 & 28.03 & 0.770 & 27.32 & 0.730  & 25.18 & 0.750  & 28.82 & 0.886 \\
  & DRRN~\cite{he2016deep}  & 31.68 & 0.889  & 28.21 & 0.772 & 27.38 & 0.728  & 25.44 & 0.764  & 29.18 & 0.891\\
  & MemNet~\cite{tai2017memnet} &31.74 & 0.889 & 28.26 & 0.772 & 27.40 & 0.728 & 25.50 & 0.763 & 29.42 & 0.894\\
  & SRMDNF~\cite{zhang2018learning} & 31.96& 0.892& 28.35& 0.778 & 27.49 &0.734& 25.68& 0.773 & 30.09 & 0.902\\
  & ASDN(ours) & \color{blue}{32.27}& \color{blue}{0.896} &\color{blue}{28.66} & \color{blue}{0.784} & \color{blue}{27.65}& \color{blue}{0.740}& \color{blue}{26.27} & \color{blue}{0.792} & \color{blue}{30.91} &\color{blue}{0.913} \\\cline{2-12}
  & LapSRN~\cite{lai2017deep}  & 31.54 & 0.885  & 28.19 & 0.772 &27.32 &0.727  & 25.21 & 0.756  & 29.46 & 0.890\\
  & EDSR ~\cite{lim2017enhanced}  & 32.46 & 0.896 & 28.80 & 0.788 &27.71 & 0.742 & 26.64 & 0.803 & 31.02& 0.915  \\
  & RDN~\cite{zhang2018residual}  & 32.47 & 0.899 & 28.81 & 0.787 &27.72& 0.742 & 26.61 & 0.803 & 31.00& 0.915   \\
  & DBPN~\cite{haris2018deep}  & 32.42 & 0.898 & 28.76 & 0.786 &27.68 & 0.740 & 26.38 & 0.796& 30.91& 0.914  \\
  & RCAN \cite{zhang2018image}  & 32.63 &  0.900 & 28.87 &0.789 & 27.77 & 0.744 & \color{red}{26.82} &\color{red}{0.809} & 31.22& 0.917  \\
  &FSDN(ours) &  \color{red}{32.63}& \color{red}{0.900} &\color{red}{28.89}&\color{red}{0.789} &\color{red}{27.79}&\color{red}{0.744}& 26.79 & 0.807& \color{red}{31.44}&\color{red}{0.919} \\\hline
  \multirow{11}{*}{\textbf{$8\times$}}& Bicubic & 24.40 & 0.658& 23.10 & 0.566  & 23.97 & 0.548  & 20.74 & 0.516  & 21.47 & 0.650 \\
  & SRCNN \cite{dong2014learning} & 25.33 & 0.690 & 23.76 & 0.591  & 24.13 & 0.566  & 21.29 & 0.544  & 22.46 & 0.695 \\
  & VDSR~\cite{kim2016accurate} & 25.93 & 0.724 & 24.26 & 0.614 & 24.49 & 0.583  & 21.70 & 0.571  & 23.16 & 0.725 \\
  & LapSRN~\cite{lai2017deep} & 26.15 & 0.738  & 24.35 & 0.620 &24.54 &0.586 & 21.81 & 0.581  & 23.39 & 0.735\\
  & MemNet~\cite{tai2017memnet}  & 26.16 & 0.741 & 24.38 & 0.619 & 24.58 & 0.584 & 21.89 & 0.583 & 23.56 & 0.738\\
  & ASDN(ours) & \color{blue}{27.02}& \color{blue}{0.776} & \color{blue}{24.99} & \color{blue}{0.641} & \color{blue}{24.82}& \color{blue}{0.600}& \color{blue}{22.57} & \color{blue}{0.620} & \color{blue}{24.73} &\color{blue}{0.748}\\\cline{2-12}
  & EDSR ~\cite{lim2017enhanced} & 26.96 &0.776 & 24.91 & 0.642& 24.81 & 0.599 & 22.51 & 0.622& 24.69& 0.784  \\
  & DBPN~\cite{haris2018deep} &27.21 &0.784 & 25.13 & 0.648 &24.88 & 0.601 & 22.73 & 0.631&25.14& 0.799  \\
  & RCAN \cite{zhang2018image} &27.31 & 0.788 & 25.23 & 0.651&24.98 & \color{red}{0.606} &\color{red}{23.00}& \color{red}{0.645}& 25.24& 0.803 \\
  & FSDN(ours)  &\color{red}{27.33}& \color{red}{0.789}&\color{red}{25.24}& \color{red}{0.651}& \color{red}{24.98}&  0.604 & 22.90& 0.638&\color{red}{25.24}&\color{red}{0.803}\\
  \hline
\end{tabular}
}
}
\end{table*}

\textbf{Testing details} Our proposed networks are tested on five widely-used benchmark datasets for image SR: Set5 \cite{bevilacqua2012low}, Set14 \cite{zeyde2010single}, BSD100 \cite{timofte2014a+}, Urban100 \cite{huang2015single} and Manga109 \cite{matsui2017sketch}. To test any-scale network (ASDN) for SR of a random scale $s$, the testing images are first downscaled with the scale factor $s$ as the LR images. If the scale $s$ is not larger than $2$, the LR images with scale $s$ are upsampled and forwarded into the ASDN with the two enabled neighboring Laplacian pyramid levels of the scale $s$. HR images are predicted by interpolating these two levels based on Eq. {\ref{2}}.
While if the scale $s$ is larger than $2$, the testing recursion times are based on $N=\ceil{\log_2 R}$. At each recursion $n$, the outputs of previous recursion are upscaled as input and deployed through ASDN with $r_{n}$ according to the Eq. {\ref{5}}, except the initial recursion, which uses the LR images as input.
To test fixed-scale network (FSDN), the testing input images are downscaled by the fixed scales $s$ and deployed into the FSDN with the scale corresponding modules are enabled to yield the testing output.

\subsection{Comparison with State-of-arts }
\begin{table*}[t]
\renewcommand{\arraystretch}{0.9}
\tiny
\caption{Results of any-scale SR on different methods tested on BSD100.  The first row shows the results of Laplacian pyramid levels and the second row demonstrates SR performance on randomly selected scales and boundary condition. \color{red}{Red} \color{black}{indicates the best performance, and} \color{blue}{blue} \color{black}{indicates the second best performance.}
}
\label{tab4}
\centering
\makebox[1\textwidth][c]{%
\resizebox{0.95\textwidth}{!}{
\begin{tabular}{|c|c|c|c|c|c|c|c|c|c|}
\hline
  \backslashbox{Method}{Scale}& \textbf{X1.1} & \textbf{X1.2}& \textbf{X1.3}& \textbf{X1.4}& \textbf{X1.5}& \textbf{X1.6}& \textbf{X1.7}& \textbf{X1.8} & \textbf{X1.9}\\
  \hline
  Bicubic & 36.56 & 35.01  & 33.84 & 32.93  & 32.14 & 31.49 & 30.90 & 30.38  & 29.97  \\
  VDSR-Conv &42.13 & 39.52  & 37.88  &36.53  & 35.42 & 34.50 & 33.72 & 33.03&32.41  \\
  EDSR-Conv &\color{blue}{42.92} & 40.11  &\color{blue}{38.33} & 36.93 & 35.79 & 34.85  & 34.06 & 33.38 & 32.75   \\
  RDN-Conv &42.86 & 40.04  & 38.25 & 36.86  & 35.72 & 34.78  & 33.99 & 33.29  & 32.67  \\
  Meta-EDSR~\cite{hu2019meta} &42.72 & 39.92 & 38.16& 36.84& 35.78 & 34.83 &34.06 &33.36& \color{blue}{32.78}\\
  Meta-RDN~\cite{hu2019meta} & 42.82& \color{red}{40.40} & 38.28 & \color{blue}{36.95}& \color{blue}{35.86} & \color{blue}{34.90}& \color{blue}{34.13} & \color{blue}{33.45} & \color{red}{32.86}\\
  ASDN(ours) &\color{red}{43.05} & \color{blue}{40.24} & \color{red}{38.42} &\color{red}{37.02} & \color{red}{35.87} & \color{red}{34.92} & \color{red}{34.14}& \color{red}{33.46} & \color{red}{32.86} \\\hline\hline
  \backslashbox{Method}{Scale}& \textbf{X2.0} & \textbf{X2.8}& \textbf{X4.0}& \textbf{X5.7}& \textbf{X8.0}& \textbf{X11.3}& \textbf{X16.0}& \textbf{X22.6} & \textbf{X32.0}\\\hline
  Bicubic & 29.55 & 27.53  & 25.96 & 24.96  & 23.67 & 22.65  & 21.73 & 20.73  & 19.90 \\
  VDSR-Conv &31.89 &29.23  & 27.25 & 25.56  & 24.58  & 23.49 & 22.47  & 21.39  & 20.38 \\
  EDSR-Conv & 32.23 & 29.54  & 27.58 & \color{blue}{26.01} & \color{blue}{24.78} & \color{blue}{23.65} &\color{blue}{22.63}& \color{blue}{21.55}&\color{blue}{20.53}\\
  RDN-Conv &32.07 & 29.47  & 27.51  & 25.94  & 24.72 & 23.60  & 22.58 & 21.50  & 20.51  \\
  Meta-EDSR~\cite{hu2019meta} &32.26 & 29.61  & \color{blue}{27.67} & -  & - & -  & - & - & - \\
  Meta-RDN~\cite{hu2019meta} &\color{red}{32.35}& \color{red}{29.67}& \color{red}{27.75} &-  & - & -  & - & - & - \\
  ASDN(ours) &\color{blue}{32.30}& \color{blue}{29.63}& 27.65 &\color{red}{26.07} &\color{red}{24.85} & \color{red}{23.70} & \color{red}{22.66} &\color{red}{21.59}& \color{red}{20.55} \\\hline
\end{tabular}
}
}
\end{table*}
To confirm the ability of the proposed methods, We first compare with state-of-the-art SR algorithms for qualitative and quantitative analysis on the normal fixed scales $2\times$, $3\times$, $4\times$, $8\times$, which includes predefined upsampling methods (SRCNN~\cite{dong2014learning}, VDSR~\cite{kim2016accurate}, DRRN~\cite{he2016deep}, MemNet~\cite{tai2017memnet} and SRMDNF~\cite{zhang2018learning}), and single upsampling methods (RDN~\cite{zhang2018residual}, LapSRN~\cite{lai2017deep}, EDSR~\cite{lim2017enhanced},  RCAN \cite{zhang2018image}).
\subsubsection{Quantitative Comparison}

We compare the performance of our any-scale SR networks with state-of-the-art methods on the five challenging dataset benchmarks. Table~\ref{tab3} shows quantitative comparisons for $\times2$, $\times3$, $\times4$, $\times8$ SR. For fair comparisons with the recent single upsampling networks, we fine-tune the ASDN with the fixed $\times2$, $\times3$, $\times4$, $\times8$ scale SR samples as FSDN for reference. It is obvious that FSDN has better performance than state-of-the-art methods, except RCAN on some datasets. Although on Urban100, which is consisted of straight-line building structure images, RCAN has better performance than FSDN due to the more channel attentions across the network, which is sensitive to the sharp edges in the image reconstruction. On other datasets, FSDN reconstruction accuracy is comparable to RCAN. This indicates the network, which is the same main framework as ASDN is effective to learn mapping functions for SR tasks.

Due to the strong ability of the framework, our ASDN performs favorably against the existing methods, especially compared to the predefined upsampling methods. Noted that ASDN does not use any $3\times$, $4\times$, $8\times$ SR samples for training but still generates comparable results as EDSR.
There are mainly two reasons for ASDN drops behind some upsampling models. First, these upsampling models are trained with fixed-scale SR samples, and customized for the $2\times$, $3\times$, $4\times$, and $8\times$ scales deployments, but ASDN is trained with scales in $(1, 2]$. Second, the upsampling layers \cite{schulter2015fast} can improve the reconstruction performance, as shown in our experiment, FSDN (the upsampling version of ASDN) has more than $0.1dB$ PSNR compared to ASDN on scale $2\times$. However, some of the upsampling layers can only apply for SR of the integer scales \cite{schulter2015fast}, such as transposed layers. Although, Meta-upsampling~\cite{hu2019meta} layer can upscale images with decimal scales, these scale factors need to be trained before deployment.
Therefore, we compromise some reconstruction accuracy for the continuous scale SR using the predefined upsampling structure, which only requires to be trained with several representative scales. Our ASDN is still very profound on the normal fixed scales compared with the existing predefined upsampling deep methods. Regarding the speed, our ASDN takes 0.5 seconds to process a 288$\times$288 image for $2\times$ SR on a Titan X GPU, and FSDN takes about 0.04 seconds to generate a 288$\times$288 image for $2\times$ SR.

\subsubsection{Any-Scale Comparison}\label{anys}

In this section, in order to evaluate the efficiency of our ASDN for any upscale ratio SR, we firstly compare ASDN with other methods.
The Bicubic interpolation method is adopted as the reference, and some deep learning network frameworks (EDSR, RDN, VDSR) are retrained with the proposed any-scale SR method and the same training data as our ASDN for any-scale SR comparison denoted as EDSR-Conv, RDN-Conv, and VDSR-Conv. Meta-EDSR and Meta-RDN~\cite{hu2019meta} are dynamic meta-upsampling models which are trained with scale factors from $\times1$ to $\times4$ at the stride of $0.1$.

\begin{figure}[h]
\begin{center}
\includegraphics[width=0.45\textwidth,height=0.20\textheight]{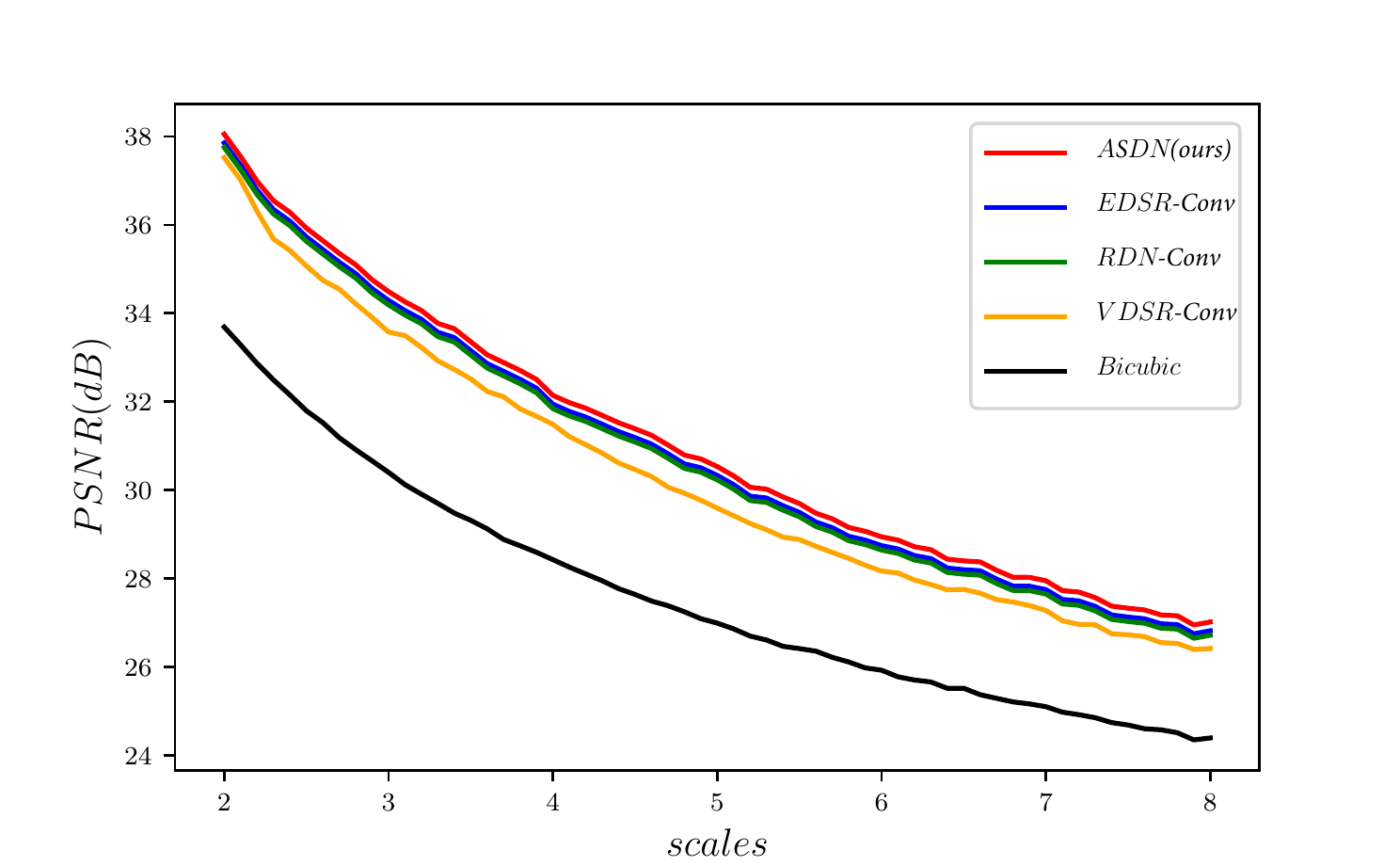}
\caption{PSNR comparison of ASDN with other works within the continuous scale range $(\times2,\times8]$ on Set5}
\label{any}
\end{center}
\end{figure}
The experimental results are shown in Table {\ref{tab4}}, which uses the PSNR value for comparison. The first row shows the PSNR value on SR of $9$ trained scales from $\times1.1$ to $\times1.9$ and it is obvious that our ASDN reaches the state-of-the-art performance.
The second row illustrates ASDN efficiency on the scales not trained before and evaluates the effective scale range of our proposed any-scale SR network. For SR of scales out of the range, ASDN is comparable to Meta-EDSR, but slightly drops behind Meta-RDN. This is due to ASDN is the recursively deployed results, and Meta-RDN is customized with these scales. Although the recursively deployed SR results have slight drop back as the directly deployed results, recursive deployment can still effectively generate SR of scales not trained before. Through this way, ASDN only needs $11$ training scales for any-scale SR.

Fig. {\ref{any}} shows the any-scale SR results on a continuous scale range. We test our any-scale network performance with random decimal scales distributed in the commonly used range of $\times2$ to $\times8$ on Set5 and plot out the results into the line.
It is proved that ASDN and the models trained with our any-scale SR method can effectively reconstruct HR images of continuous upscale ratios. Our ASDN outperforms all the other methods, which is generally $0.15$ dB better than EDSR-Conv, outperforms VDSR-Conv by 0.6 dB and robustly keeps the deference of more than $3$ dB PSNR from Bicubic method in the continuous scale range. The result demonstrates our ASDN can effectively reconstruct HR images of continuous upscale ratios and our any-scale training method is flexible to many deep CNN networks.
\begin{figure*}[htbp]
\begin{center}
\includegraphics[width=0.90\textwidth,height=0.20\textheight]{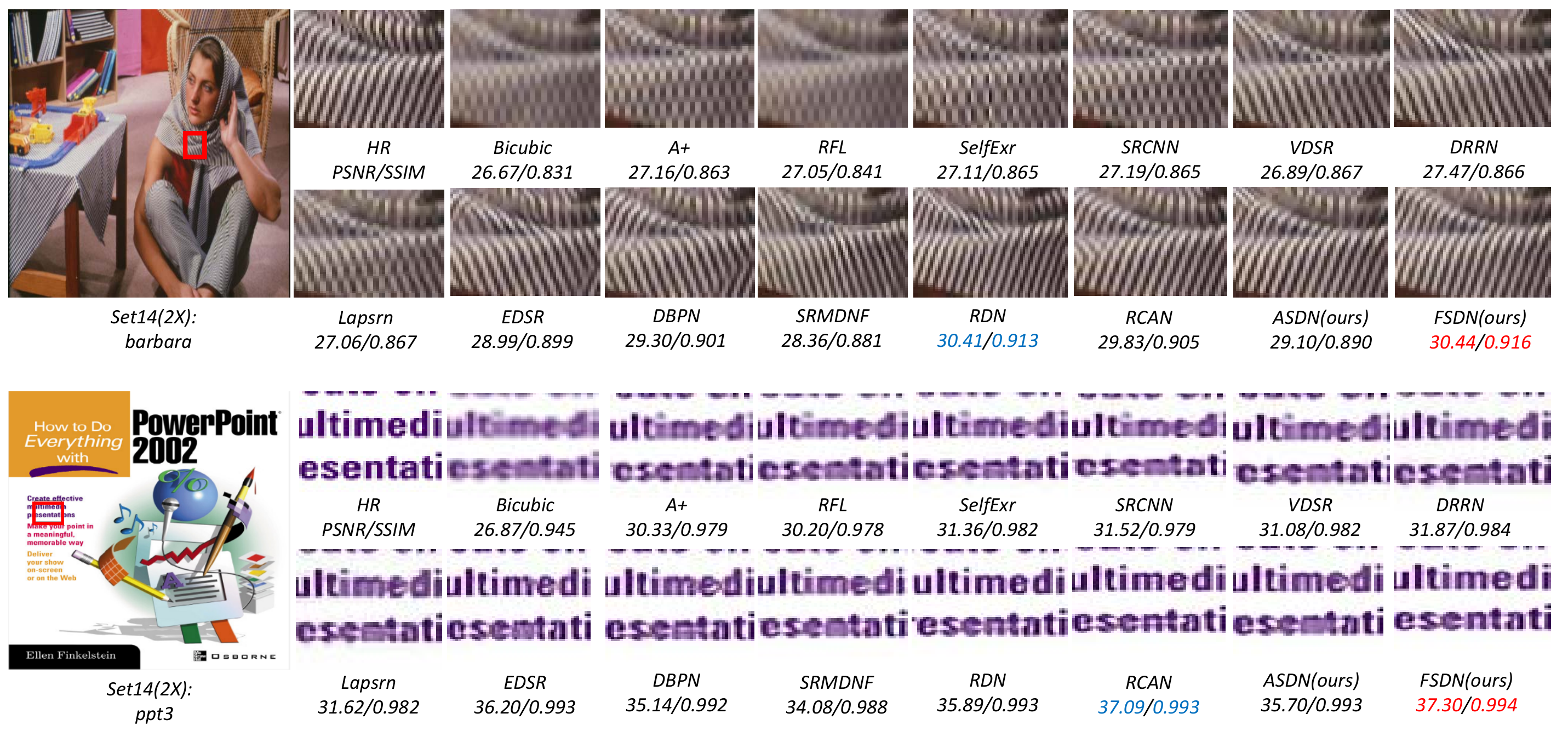}
\caption{Qualitative comparisons of our models with other works on $\times2 $ super-resolution. \color{red}{Red} \color{black}{indicates the best performance, and} \color{blue}{blue} \color{black}{indicates the second best} }
\label{fig2}
\end{center}
\end{figure*}

\begin{figure*}[htbp]
\begin{center}
\includegraphics[width=0.90\textwidth,height=0.20\textheight]{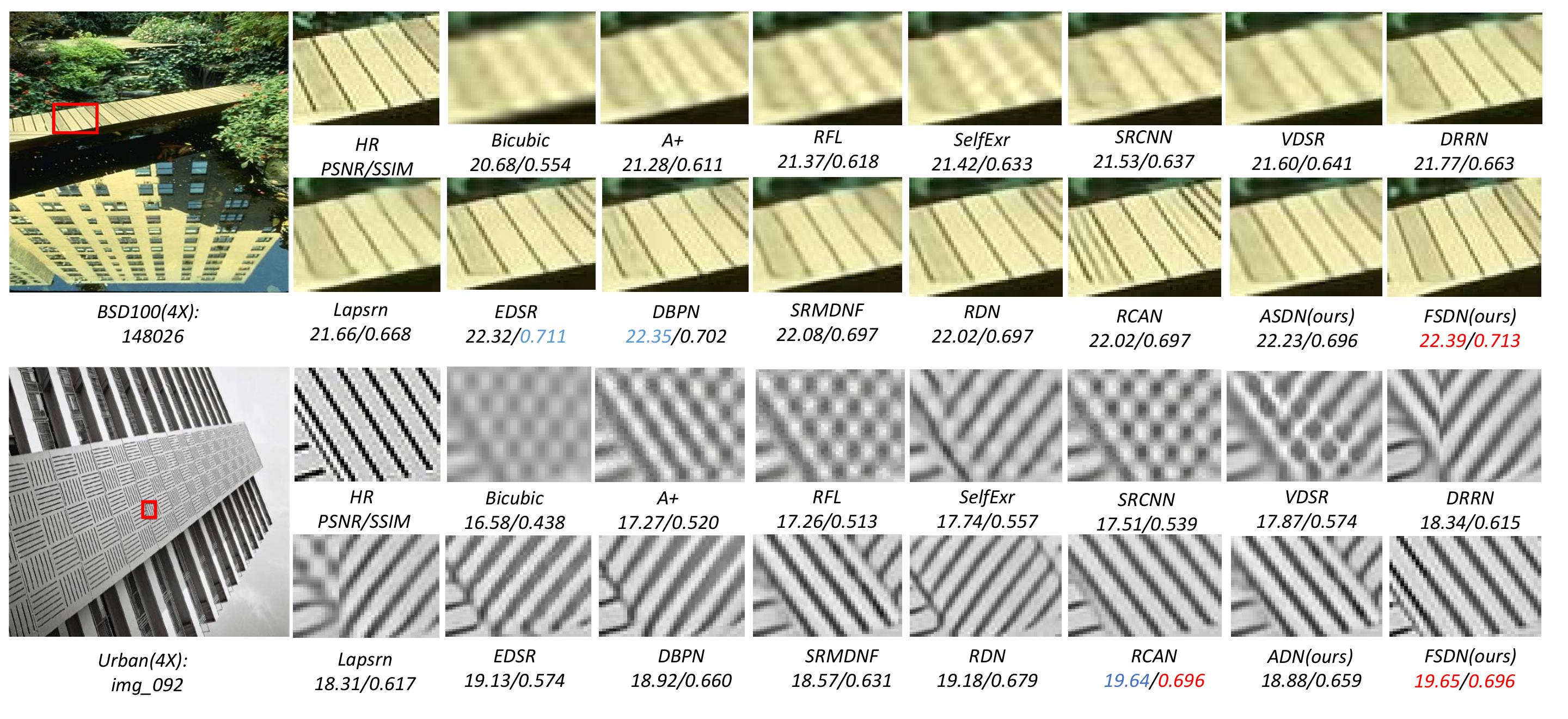}
\caption{Qualitative comparisons of our models with other works on $\times4 $ super-resolution. \color{red}{Red} \color{black}{indicates the best performance, and} \color{blue}{blue} \color{black}{indicates the second best}}
\label{fig4}
\end{center}
\end{figure*}

\begin{figure*}[htbp]
\begin{center}
\includegraphics[width=0.90\textwidth,height=0.20\textheight]{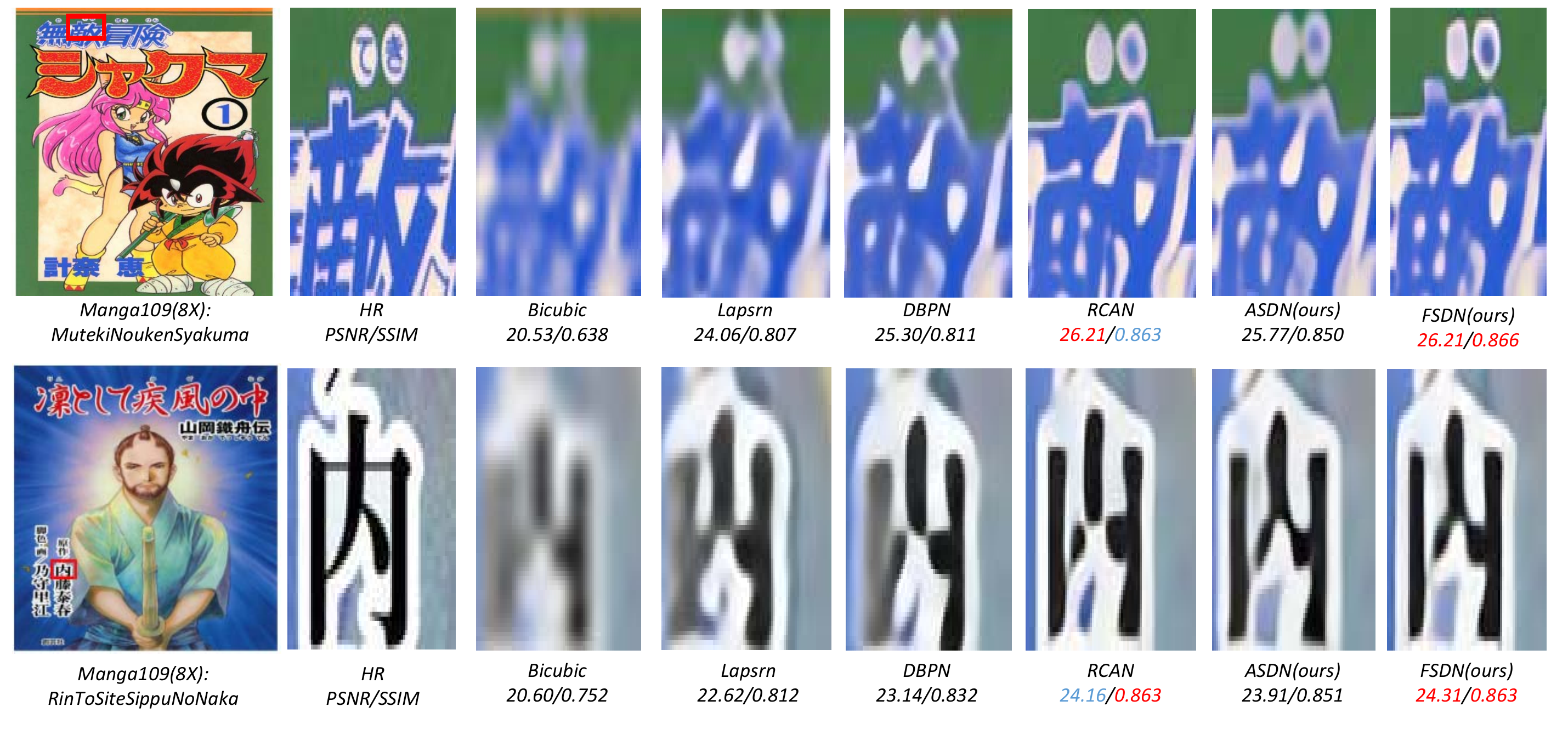}
\caption{Qualitative comparisons of our models with other works on $\times8 $ super-resolution. \color{red}{Red} \color{black}{indicates the best performance, and} \color{blue}{blue} \color{black}{indicates the second best} }
\label{fig8}
\end{center}
\end{figure*}

\begin{figure*}[t]%
    \centering
    \subfloat[Study of Effectiveness]{{\includegraphics[width=0.45\textwidth,height=0.20\textheight]{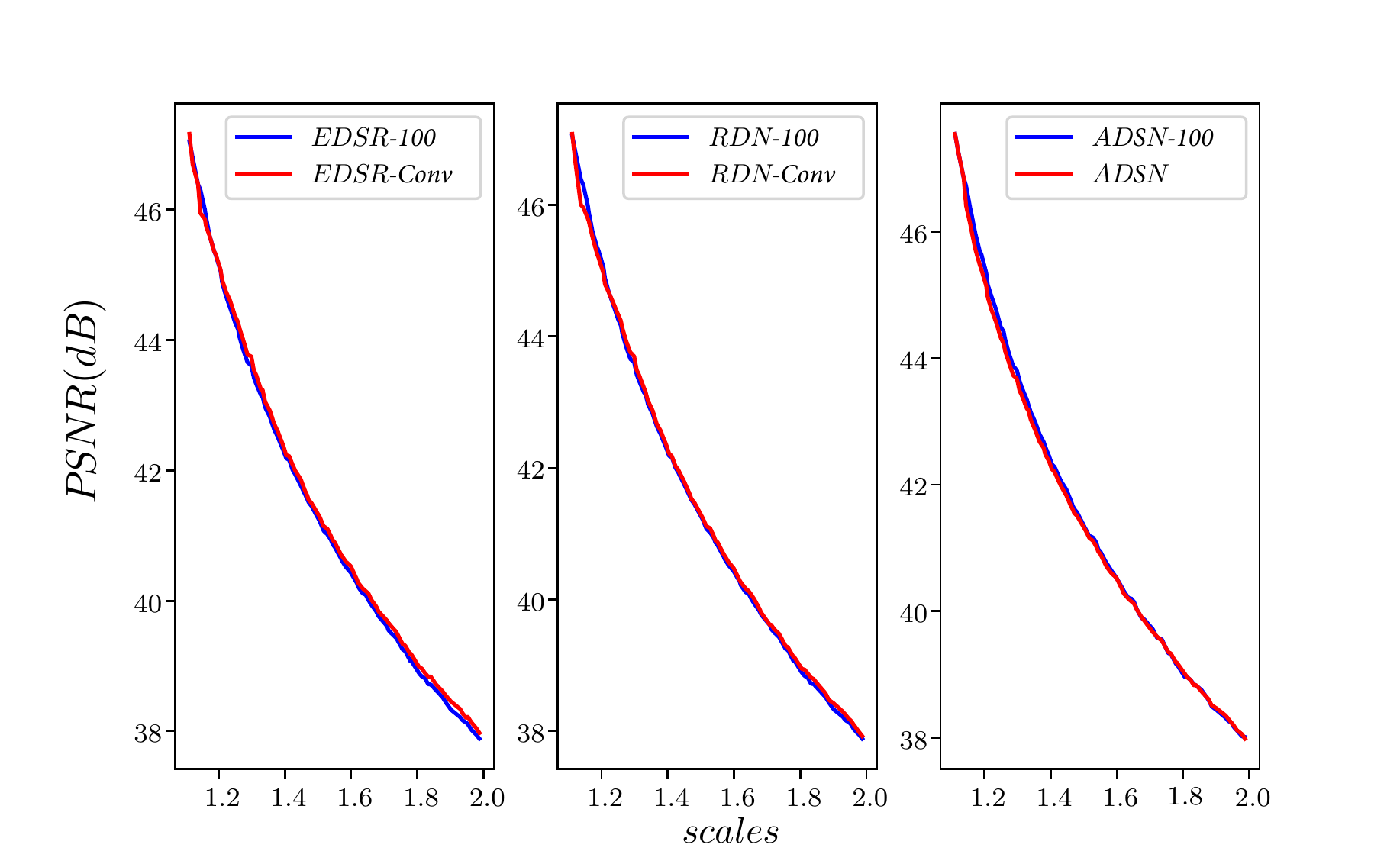} }}%
    \qquad
    \subfloat[Study of Density]{{\includegraphics[width=0.45\textwidth,height=0.20\textheight]{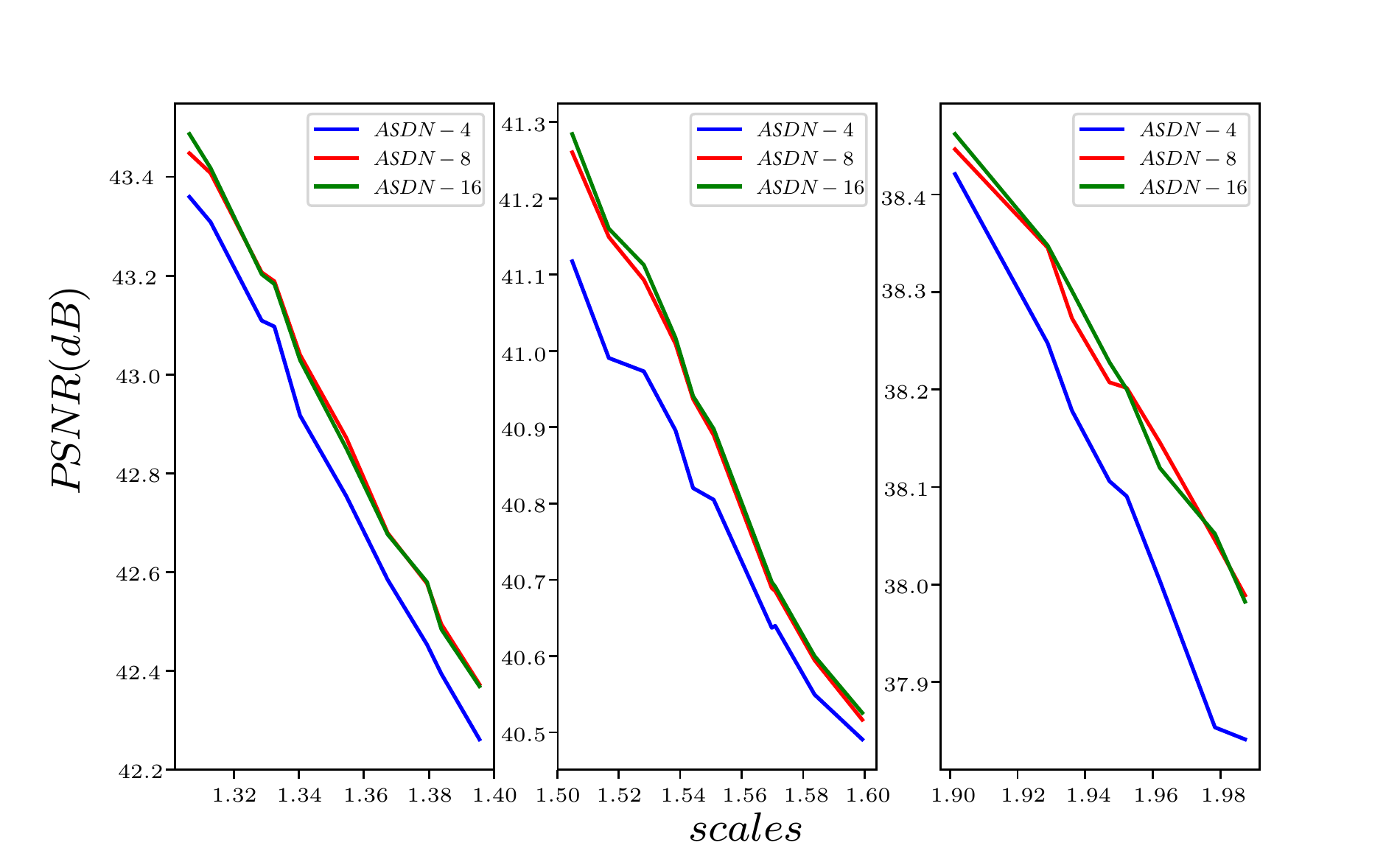} }}%
    \caption{Study of Laplacian Frequency Representation}%
    \label{fig:example}%
\end{figure*}

\subsubsection{Qualitative Comparison}
We show visual comparisons on the testing datasets for $2\times$, $4\times$ and $8\times$ SR.
For $2\times$ enlargement of Set14 in Fig. {\ref{fig2}}, FSDN suppresses the bias artifacts and recovers the cloth pattern and text closer to the ground truths than all the other methods. Meanwhile, ASDN tends to construct less biased images than other methods. For $4\times$ enlargement of the parallel straight lines in Fig. {\ref{fig4}}. Our methods generate a clearer building line, while other methods suffer the blurring artifacts. RCAN tends to generate misleading strong edges due to the more channel attention structure, but our ASDN and FSDN generates soft patterns closer to the ground truth.
The reconstruction performance on $8\times$ SR is further analyzed in Fig. {\ref{fig8}}. FSDN restores the sharper characters than the compared networks and ASDN is able to recover more accurate textures from the distorted LR image than many other fixed-scale methods.

\subsection{Study of Any-Scale Methods}
We study the effects of Laplacian Frequency Representation and Recursive Deployment of the any-scale SR methods.

\subsubsection{Laplacian Frequency Representation}\label{density}
To evaluate the accuracy of the Laplacian Frequency Representation for continuous scale SR. We compare the reconstruction results of the Laplacian Frequency Representation with the directly deployed HR images of $100$ scales in the range $(1,2]$.

We first modify EDSR, RDN, and ASDN frameworks into the single predefined upsampling networks and train them with these 100 scales SR samples as EDSR-100, RDN-100 and ASDN-100 to generate HR images of Set5 on the $100$ scales. Then we reconstruct the single redefined upsampling EDSR-100 and RDN-100 with $11$ parallel IRBs as EDSR-Conv and RDN-Conv, as suggested in Sec. \ref{anys}, trained with the same method and data as ASDN.
As shown in Fig. {\ref{fig:example}}(a), It is obvious that the Laplacian frequency represented HR images have a similar quality to the direct deployed HR images.

To analyze the influence of the Laplacian pyramid level density on the SR performance, we train ASDN on $5, 9, 17$ evenly distributed upscale decimal ratios in $(1,2]$ with DIV2K, which separates the Laplacian Frequency Representation into $4, 8$ and $16$ phases and names ASDN-4, ASDN-8, and ASDN-16 separately.
Fig. {\ref{fig:example}}(b) demonstrates the performance of the three versions of ASDN with scales in $(1,2]$. In order to make the difference more obvious, we choose some scale ranges in $(1,2]$. It illustrates that ASDN-4 drops behind ASDN-8 and ASDN-16 commonly, and ASDN-8 and ASDN-16 almost overlap. The results show the Laplacian pyramid level density influences SR performance. To some extent, the model trained with more dense scales achieves better performance, but it saturates beyond a certain point, such as $10$ phases. Due to this reason, we can generate HR images of any decimal scale in the range of $(1,2]$ by the several Laplacian pyramid levels in $(1,2]$.


\subsubsection{Recursive Deployment}\label{rec}
In order to investigate the effects of recursive deployment for HR images of larger decimal scales.
We mainly demonstrate the comparison of recursive deployment and direct deployment on scales $\times2, \times3, \times4$
We trained VDSR-Conv, EDSR-Conv, RDN-Conv, and ASDN with $11$ evenly distributed upscale decimal ratios in $(1,2]$ as the recursive models and the HR images are twice upscaled with the upscale ratios $\times\sqrt{2}$, $\times\sqrt{3}$, $\times\sqrt{4}$. To form the fair comparisons, we trained VDSR-Conv, EDSR-Conv, RDN-Conv, and ASDN with $\times2, \times3, \times4$ SR images as the direct deployment models.
Table \ref{tab1} illustrates the PSNR of recursive deployment and direct deployment. It is obvious that recursive deployment generally leads to the SR performance decline compared to the direct deployment. But the difference between recursive deployment and direct deployment goes down as the scale goes up. Since the decline is still in an acceptable range and goes gentle as the upscale ratios up, we adopt recursive deployment for SR in higher upscale ratio ranges.
\begin{table}[htbp]
\tiny
\begin{center}
\resizebox{0.48\textwidth}{!}{
\begin{tabular}{|c|c|c|c|c|c|c|}
\hline
 Methods & \multicolumn{3}{|c|}{Direct deployment} & \multicolumn{3}{|c|}{ Recursive deployment} \\
 \cline{2-7}
 & \textbf{$\times2$} &\textbf{$\times3$} &\textbf{$\times4$} &\textbf{$\times2$} &\textbf{$\times3$} &\textbf{$\times4$}\\
\hline
VDSR-Conv & 37.57& 33.77 & 31.56 & 36.86 & 33.70 & 31.50   \\
\hline
EDSR-Conv & 38.04 & 34.45  & 32.29 & 37.18  & 34.32 & 32.26 \\
\hline
RDN-Conv & 38.05 & 34.46  & 32.31 & 37.27 & 34.38 & 32.23\\
\hline
Ours & 38.12& 34.52 & 32.28 & 37.35  & 34.43 & 32.27  \\
\hline
\end{tabular}
}
\end{center}
\caption{PSNR of the recursive deployment and direct deployment on SR for $\times2, \times3, \times4$}
\label{tab1}
\end{table}

To determine the best solution of recursive times $N$ and upscale ratios $r$ for recursive deployment.
We also explore various combinations of $N$ and $r$ to deploy any-scale HR images with different strategies. Table \ref{tab2} illustrates the performance of the HR images deployed by different strategies with ASDN on Set5. It is obvious that the larger upscale ratio $r$ combined with, the smaller recursive time $N$ will contribute to better performance. Furthermore, choosing the larger upscale ratios in the early recursions can produce better results than using the smaller scales. For these reasons, we recommend choosing $N=\ceil{\log_2 R}$ with the largest upscale ratios $r=2$ at the early $N-1_{th}$ recursions for large scale SR.
\begin{table}[htbp]
\tiny
\resizebox{0.48\textwidth}{!}{
\begin{tabular}{|c|c|c|c|}
\hline
Scale(R) & Recursion(N) & UpscaleRatio(r) & PSNR \\
\hline
\multirow{4}{*}{\textbf{$3\times$}}
& \multirow{3}{*}{2} & 1.732, 1.732 & 34.43\\
\cline{3-4}
&  & 1.500, 2.000 & 34.19 \\
\cline{3-4}
&  & 2.000, 1.500 & \textbf{34.48} \\
\cline{2-4}
 & 3 & 1.442, 1.442, 1.442 & 33.18\\
\hline
\multirow{4}{*}{\textbf{$4\times$}} & 2 & 2.000, 2.000  &\textbf{32.27}  \\
\cline{2-4}
& \multirow{3}{*}{3} & 1.587, 1.587, 1.587 & 31.96 \\
\cline{3-4}
&  & 1.800, 1.800, 1.234 & 32.16 \\
\cline{3-4}
&  & 2.000, 1.800, 1.100 &  32.24\\
\hline
\end{tabular}%
}
\caption{PSNR of recursive deployment and direct deployment on SR for $\times2, \times3, \times4$. \textbf{Black} indicates the best performance}
\label{tab2}
\end{table}

\subsection{Model Analysis}

\subsubsection{Number of parameters}
\begin{figure}[htbp]
\begin{center}
\includegraphics[width=0.45\textwidth,height=0.18\textheight]{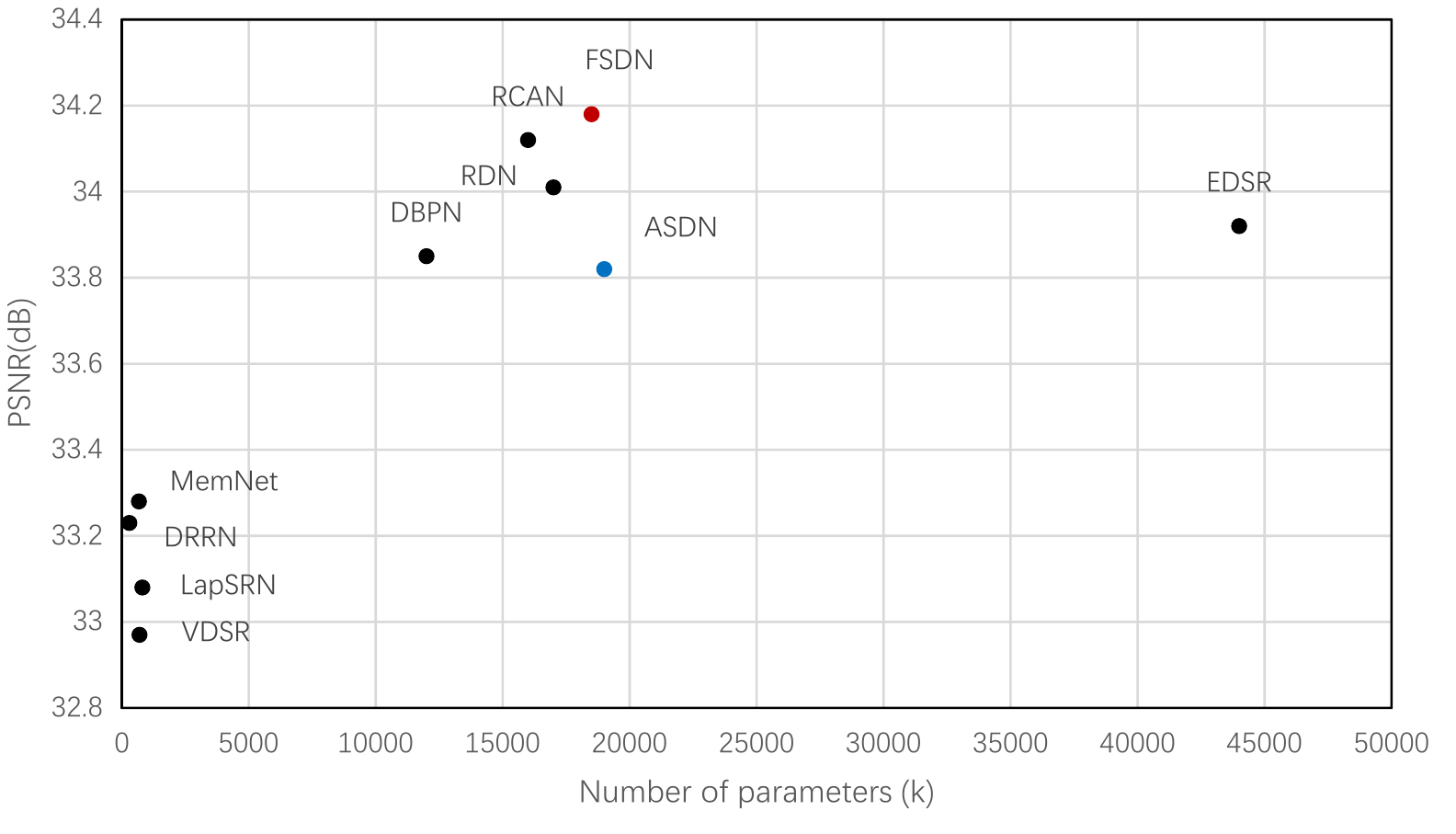}
\caption{Performance vs number of parameters. The results are evaluated with Set14 for $2\times$ enlargement. \color{red}{Red} \color{black}{indicates the best performance, and} \color{blue}{blue} \color{black}{indicates the best performance among predefined upsampling methods}}
\label{compare}
\end{center}
\end{figure}

To demonstrate the compactness of our model, we compare the model performance and network parameters of our model with the existing deep networks for image SR in Fig. {\ref{compare}}. Our model shows the trade-off between the parameter demands and performance.
Since VDSR, DRRN, LapSRN, and MemNet are all light version networks, they all visibly concede the performance for the model parameter numbers. Therefore ASDN outperforms all the other predefined upsampling methods over 0.5 dB on Set14 for $2\times$ enlargement. Furthermore, FSDN achieves the best results with a moderate number of parameters compared to all the other upsampling methods.
\subsubsection{Abviation Study}\label{abv}

In this section, we evaluate the influence of different network modules, such as channel attention (CA) in FMB, spatial attention (SA) in IRB, and skip connection (SC) between input and output.
To demonstrate the effect of CA in the proposed network structure, we remove the CA from the FMB. In Table \ref{attention}, we can see when CA is removed, the PSNR value on Set5 $(\times2)$ is relatively low compared to the model having CA.
To investigate the effect of SA, we remove the SA from the ASDN to compare with the network with SA. SA can improve performance by 0.02 dB or 0.01 dB with or without CA in the models.
We further investigate the contribution of SC to the network by comparing the models with or without SC. Adding global skip connections between the network input and output generally improves 0.04 dB on Set5. Generally combining attention modules into the network design, helps the residual high-frequency information reconstruction.
\begin{table}[htbp]
\tiny
\begin{center}
\resizebox{0.48\textwidth}{!}{
\begin{tabular}{|c|c|c|c|c|c|c|c|c|}
\hline
Module  & \multicolumn{8}{c|}{Different combination of CA, SA and SC}\\
\hline
CA & $\times $ & $\times $ & $\times $ & $\surd $ & $\surd $ & $\surd $ & $\times $ & $\surd $\\
\hline
SA & $\times $ & $\times $ & $\surd  $ & $\times$ & $\surd $ & $\times $ & $\surd $& $\surd $ \\
\hline
SC & $\times $ & $\surd $ & $\times $ & $\times $ & $\times $ & $\surd $ & $\surd $& $\surd $\\
\hline
PSNR& 37.92 & 37.96 & 37.93 & 37.95 & 37.97 & 37.99 & 37.97 & \textbf{38.01}\\
\hline
\end{tabular}
}
\end{center}
\caption{Investigation of channel attention (CA), spatial attention (SA), and skip connection (SC). \textbf{Black} indicates the best performance}
\label{attention}
\end{table}
\section{Conclusion}
In this paper, we propose an any-scale deep network (ASDN) to generate HR images of any scale with one unified network by adopting our proposed any-scale SR method, including Laplacian Frequency Representation for SR of small continuous scale ranges and Recursive Deployment for larger-scale SR. The any-scale SR method helps to reduce the demands of training scale samples and accelerate the network convergence. The extensive comparisons show our ASDN is superior to the most state-of-the-art methods on both fixed-scale and any-scale benchmarks.


%
%

{
\small
\bibliographystyle{spmpsci}      
\bibliography{article}   

\begin{thebibliography}{10}
\providecommand{\url}[1]{{#1}}
\providecommand{\urlprefix}{URL }
\expandafter\ifx\csname urlstyle\endcsname\relax
  \providecommand{\doi}[1]{DOI~\discretionary{}{}{}#1}\else
  \providecommand{\doi}{DOI~\discretionary{}{}{}\begingroup
  \urlstyle{rm}\Url}\fi

\bibitem{agustsson2017ntire}
Agustsson, E., Timofte, R.: Ntire 2017 challenge on single image
  super-resolution: Dataset and study.
\newblock In: The IEEE Conference on Computer Vision and Pattern Recognition
  (CVPR) Workshops, vol.~3, p.~2 (2017)

\bibitem{bevilacqua2012low}
Bevilacqua, M., Roumy, A., Guillemot, C., Alberi-Morel, M.L.: Low-complexity
  single-image super-resolution based on nonnegative neighbor embedding  (2012)

\bibitem{burt1987laplacian}
Burt, P.J., Adelson, E.H.: The laplacian pyramid as a compact image code.
\newblock In: Readings in Computer Vision, pp. 671--679. Elsevier (1987)

\bibitem{dong2014learning}
Dong, C., Loy, C.C., He, K., Tang, X.: Learning a deep convolutional network
  for image super-resolution.
\newblock In: European Conference on Computer Vision, pp. 184--199. Springer
  (2014)

\bibitem{gao2018face}
Gao, G., Zhu, D., Yang, M., Lu, H., Yang, W., Gao, H.: Face image
  super-resolution with pose via nuclear norm regularized structural orthogonal
  procrustes regression.
\newblock Neural Computing and Applications pp. 1--11 (2018)

\bibitem{haris2018deep}
Haris, M., Shakhnarovich, G., Ukita, N.: Deep backprojection networks for
  super-resolution.
\newblock In: Conference on Computer Vision and Pattern Recognition (2018)

\bibitem{he2016deep}
He, K., Zhang, X., Ren, S., Sun, J.: Deep residual learning for image
  recognition.
\newblock In: Proceedings of the IEEE conference on computer vision and pattern
  recognition, pp. 770--778 (2016)

\bibitem{hu2019meta}
Hu, X., Mu, H., Zhang, X., Wang, Z., Tan, T., Sun, J.: Meta-sr: a
  magnification-arbitrary network for super-resolution.
\newblock In: Proceedings of the IEEE Conference on Computer Vision and Pattern
  Recognition, pp. 1575--1584 (2019)

\bibitem{hu2019channel}
Hu, Y., Li, J., Huang, Y., Gao, X.: Channel-wise and spatial feature modulation
  network for single image super-resolution.
\newblock IEEE Transactions on Circuits and Systems for Video Technology
  (2019)

\bibitem{huang2016deep}
Huang, G., Sun, Y., Liu, Z., Sedra, D., Weinberger, K.Q.: Deep networks with
  stochastic depth.
\newblock In: European Conference on Computer Vision, pp. 646--661. Springer
  (2016)

\bibitem{huang2015single}
Huang, J.B., Singh, A., Ahuja, N.: Single image super-resolution from
  transformed self-exemplars.
\newblock In: Proceedings of the IEEE Conference on Computer Vision and Pattern
  Recognition, pp. 5197--5206 (2015)

\bibitem{kim2016accurate}
Kim, J., Kwon~Lee, J., Mu~Lee, K.: Accurate image super-resolution using very
  deep convolutional networks.
\newblock In: Proceedings of the IEEE Conference on Computer Vision and Pattern
  Recognition, pp. 1646--1654 (2016)

\bibitem{lai2017deep}
Lai, W.S., Huang, J.B., Ahuja, N., Yang, M.H.: Deep laplacian pyramid networks
  for fast and accurate superresolution.
\newblock In: IEEE Conference on Computer Vision and Pattern Recognition,
  vol.~2, p.~5 (2017)

\bibitem{lai2018fast}
Lai, W.S., Huang, J.B., Ahuja, N., Yang, M.H.: Fast and accurate image
  super-resolution with deep laplacian pyramid networks.
\newblock IEEE transactions on pattern analysis and machine intelligence
  (2018)

\bibitem{lim2017enhanced}
Lim, B., Son, S., Kim, H., Nah, S., Lee, K.M.: Enhanced deep residual networks
  for single image super-resolution.
\newblock In: The IEEE conference on computer vision and pattern recognition
  (CVPR) workshops, vol.~1, p.~4 (2017)

\bibitem{liu2018attention}
Liu, Y., Wang, Y., Li, N., Cheng, X., Zhang, Y., Huang, Y., Lu, G.: An
  attention-based approach for single image super resolution.
\newblock In: 2018 24th International Conference on Pattern Recognition (ICPR),
  pp. 2777--2784. IEEE (2018)

\bibitem{lu2017underwater}
Lu, H., Li, Y., Nakashima, S., Kim, H., Serikawa, S.: Underwater image
  super-resolution by descattering and fusion.
\newblock IEEE Access \textbf{5}, 670--679 (2017)

\bibitem{lu2019satellite}
Lu, T., Wang, J., Zhang, Y., Wang, Z., Jiang, J.: Satellite image
  super-resolution via multi-scale residual deep neural network.
\newblock Remote Sensing \textbf{11}(13), 1588 (2019)

\bibitem{matsui2017sketch}
Matsui, Y., Ito, K., Aramaki, Y., Fujimoto, A., Ogawa, T., Yamasaki, T.,
  Aizawa, K.: Sketch-based manga retrieval using manga109 dataset.
\newblock Multimedia Tools and Applications \textbf{76}(20), 21811--21838
  (2017)

\bibitem{schulter2015fast}
Schulter, S., Leistner, C., Bischof, H.: Fast and accurate image upscaling with
  super-resolution forests.
\newblock In: Proceedings of the IEEE Conference on Computer Vision and Pattern
  Recognition, pp. 3791--3799 (2015)

\bibitem{tai2017memnet}
Tai, Y., Yang, J., Liu, X., Xu, C.: Memnet: A persistent memory network for
  image restoration.
\newblock In: Proceedings of the IEEE Conference on Computer Vision and Pattern
  Recognition, pp. 4539--4547 (2017)

\bibitem{timofte2014a+}
Timofte, R., De~Smet, V., Van~Gool, L.: A+: Adjusted anchored neighborhood
  regression for fast super-resolution.
\newblock In: Asian Conference on Computer Vision, pp. 111--126. Springer
  (2014)

\bibitem{tong2017image}
Tong, T., Li, G., Liu, X., Gao, Q.: Image super-resolution using dense skip
  connections.
\newblock In: Computer Vision (ICCV), 2017 IEEE International Conference on,
  pp. 4809--4817. IEEE (2017)

\bibitem{wang2015robust}
Wang, D., Lu, H., Yang, M.H.: Robust visual tracking via least soft-threshold
  squares.
\newblock IEEE Transactions on Circuits and Systems for Video Technology
  \textbf{26}(9), 1709--1721 (2015)

\bibitem{wang2018deep}
Wang, Y., Shen, J., Zhang, J.: Deep bi-dense networks for image
  super-resolution.
\newblock In: 2018 Digital Image Computing: Techniques and Applications
  (DICTA), pp. 1--8. IEEE (2018)

\bibitem{zeyde2010single}
Zeyde, R., Elad, M., Protter, M.: On single image scale-up using
  sparse-representations.
\newblock International conference on curves and surfaces pp. 711--730 (2010)

\bibitem{zhang2018learning}
Zhang, K., Zuo, W., Zhang, L.: Learning a single convolutional super-resolution
  network for multiple degradations.
\newblock In: IEEE Conference on Computer Vision and Pattern Recognition,
  vol.~6 (2018)

\bibitem{zhang2018image}
Zhang, Y., Li, K., Li, K., Wang, L., Zhong, B., Fu, Y.: Image super-resolution
  using very deep residual channel attention networks.
\newblock In: Proceedings of the European Conference on Computer Vision (ECCV),
  pp. 286--301 (2018)

\bibitem{zhang2018residual}
Zhang, Y., Tian, Y., Kong, Y., Zhong, B., Fu, Y.: Residual dense network for
  image super-resolution.
\newblock In: The IEEE Conference on Computer Vision and Pattern Recognition
  (CVPR) (2018)

\end{thebibliography}
}
%
%

\end{document}